\documentclass[iop, revtex4]{emulateapj}

\usepackage{color}
\usepackage{xcolor}
\usepackage{courier}
\usepackage{lscape}
\usepackage{sidecap}
\usepackage{hyperref}

\shortauthors{Smith et al.}
\shorttitle{TESS-Detected GRB~191016A}

\begin{document}

\title{GRB 191016A: A Long Gamma-Ray Burst Detected by TESS}

\author{Krista Lynne Smith\altaffilmark{1,2,3}}
\author{Ryan Ridden-Harper\altaffilmark{4}}
\author{Michael Fausnaugh\altaffilmark{5}}
\author{Tansu Daylan\altaffilmark{5}}
\author{Nicola Omodei\altaffilmark{6}}
\author{Judith Racusin\altaffilmark{7}}
\author{Zachary Weaver\altaffilmark{8}}
\author{Thomas Barclay\altaffilmark{9,10}}
\author{P\'eter Veres\altaffilmark{11}}
\author{D. Alexander Kann\altaffilmark{12}}
\author{Makoto Arimoto\altaffilmark{13}}

\altaffiltext{1}{KIPAC at SLAC, Stanford University, Menlo Park, CA 94025, USA; klynne@stanford.edu}
\altaffiltext{2}{Einstein Fellow}
\altaffiltext{3}{Southern Methodist University, Department of Physics, Dallas, TX 75205, USA}
\altaffiltext{4}{Research School of Astronomy \& Astrophysics, Mount Stromlo Observatory, The Australian National University, Cotter Road, Weston Creek, ACT 2611, Australia}
\altaffiltext{5}{MIT Kavli Institute for Astrophysics and Space Research, Massachusetts Institute of Technology, Cambridge, MA, USA}
\altaffiltext{6}{W. W. Hansen Experimental Physics Laboratory, Kavli Institute for Particle Astrophysics and Cosmology, Department of Physics and SLAC National Accelerator Laboratory, Stanford University, Stanford, CA 94305, USA}
\altaffiltext{7}{Astrophysics Science Division, NASA Goddard Space Flight Center, Mail Code 661, Greenbelt, MD 20771, USA}
\altaffiltext{8}{Institute for Astrophysical Research, Boston University, 725 Commonwealth Avenue, Boston, MA, 02215 USA}
\altaffiltext{9}{Exoplanets and Stellar Astrophysics Laboratory, Code 667, NASA Goddard Space Flight Center, Greenbelt, MD 20771, USA}
\altaffiltext{10}{University of Maryland, Baltimore County, 1000 Hilltop Circle, Baltimore, MD 21250, USA}
\altaffiltext{11}{Center for Space Plasma and Aeronomic Research, University of Alabama in Huntsville, Huntsville, AL 35899, USA}
\altaffiltext{12}{Instituto de Astrof\'{i}sica de Andaluc\'{i}a (IAA-CSIC), Glorieta de la Astronom\'{i}a s/n, 18008 Granada, Spain}
\altaffiltext{13}{Faculty of Mathematics and Physics, Institute of Science and Engineering, Kanazawa University, Kakuma, Kanazawa, Ishikawa 920-1192, Japan}

\begin{abstract}

The TESS exoplanet-hunting mission detected the rising and decaying optical afterglow of GRB~191016A, a long Gamma-Ray Burst (GRB) detected by \emph{Swift}-BAT but without prompt XRT or UVOT follow-up due to proximity to the moon. The afterglow has a late peak at least 1000~seconds after the BAT trigger, with a brightest-detected TESS datapoint at 2589.7~s post-trigger. The burst was not detected by \emph{Fermi}-LAT, but was detected by \emph{Fermi}-GBM without triggering, possibly due to the gradual nature of rising light curve. Using ground-based photometry, we estimate a photometric redshift of $z_\mathrm{phot} = 3.29\pm{0.40}$. Combined with the high-energy emission and optical peak time derived from TESS, estimates of the bulk Lorentz factor $\Gamma_\mathrm{BL}$ range from $90-133$. The burst is relatively bright, with a peak optical magnitude in ground-based follow-up of $R=15.1$~mag. Using published distributions of GRB afterglows and considering the TESS sensitivity and sampling, we estimate that TESS is likely to detect $\sim1$ GRB afterglow per year above its magnitude limit.

\end{abstract}


\section{Introduction}
\label{sec:intro}

The Transiting Exoplanet Survey Satellite, or TESS \citep{Ricker2015}, is currently in the midst of a nearly all-sky timing survey in search of transiting planets around M-dwarfs. On 2019 October 16 the Burst Alert Telescope \citep[BAT;][]{2005SSRv..120..143B} onboard the \emph{Neil Gehrels Swift Observatory} \citep[][\emph{Swift} hereafter]{2004ApJ...611.1005G} detected a gamma ray burst (GRB) in the portion of the sky being monitored by TESS \citep{Gropp2019}. The burst occurred too close to the moon for \emph{Swift} to safely slew to its position, preventing UVOT and XRT follow-up until over 11~hours after the BAT trigger. The TESS data are therefore the only space-based follow-up for the burst before this time. Several ground-based observatories detected the counterpart and afterglow simultaneously with the TESS measurements, as documented in the GRB Coordinates Network (GCN)\footnote{https://gcn.gsfc.nasa.gov/}\citep{Watson2019b,Watson2019a,Zheng2019,Hu2019,Kim2019,Schady2019,Toma2019}.

TESS clearly detects the rising pre-peak light curve of this long GRB. Here, we present the TESS light curve and discuss the viability of TESS data to help constrain properties of GRBs that happen to occur within its field of view.

In Section~\ref{sec:grb}, we give the parameters of the burst as reported by \emph{Swift}. In Section~\ref{sec:tess} we discuss the TESS mission and the extraction of the light curve. Section~\ref{sec:fermi} discusses the high energy emission from the burst as observed by \emph{Fermi}, while Section~\ref{sec:latepeak} presents the optical afterglow properties and how they compare to known bursts. Section~\ref{sec:redshift} presents the photometric modelling to determine the redshift of the afterglow.  In Section~\ref{sec:lorentz}, we use the redshift and burst parameters to calculate the bulk Lorentz factor. Section~\ref{sec:discussion} presents the calculation of how many bursts may be observable with TESS, and Section~\ref{sec:conclusion} provides a brief conclusive summary.

When redshift is dealt with in this document, we have assumed a cosmology with $H_0=69.6$~km~s$^{-1}$ Mpc$^{-1}$, $\Omega_M = 0.286$, and $\Omega_\Lambda = 0.714$.

\section{GRB 191016A}
\label{sec:grb}
The BAT trigger for GRB~191016A occurred at 04:09:00.91 UT on 2019 October 16. The enhanced position was reported as RA=02:01:04.67, DEC=+24:30:35.3 (J2000), corresponding to no cataloged galaxy. The BAT light curve lasts for approximately 210 seconds after the trigger, with a poorly-constrained T$_{90}$ duration of 
220$\pm{180}$~s. 
The burst had a fluence of 1.12$\times10^{-5}$~erg~cm$^{-2}$ in the energy range $15-350$~keV regardless of whether the light curve is modelled as a power law with or without a cutoff; in both models the best fitting power law photon index is $\Gamma_\mathrm{ph}=1.54\pm{0.09}$, where the sign convention used is $E^{-\Gamma_\mathrm{ph}}$. 

Due to the moon constraints of the \emph{Swift}-XRT and UVOT telescopes, the satellite did not immediately slew to the position of the burst. This means that the XRT and UVOT light curves of the afterglow, typically simultaneous with ground-based follow-up, are delayed by several hours. 
GRB\,191016A was at 72$^{\circ}$ from the \emph{Fermi}-Large Area Telescope \citep[LAT,][]{LATPaper} boresight. The burst entered the \emph{Fermi}-LAT field of view about 4~ks after the \emph{Swift}-BAT trigger, but was not detected by LAT. Despite being in the field-of-view of the \emph{Fermi} Gamma-ray Burst Monitor \citep[GBM, ][]{meeg09}, it did not cause a trigger but was detected; see Section~\ref{sec:fermi} for details. 

The earliest follow-up of the burst was ground-based, beginning a few hundred seconds after the trigger, as chronicled in the GCN (for references, see the Introduction); the burst peaks at an apparent magnitude of $R=15.1$~mag. 

\section{The TESS Mission and Light Curve}
\label{sec:tess}

During the recently completed TESS primary mission phase, the spacecraft observed each 24$\times$96 degree sector of the sky (over 2300 square degrees) continuously for ~27 days, recording integrations every 30-minutes, before moving on to the next sector. At high ecliptic latitudes, the sectors overlap, resulting in light curves with extended durations. The bandpass is wide and monolithic, spanning the red-optical to the NIR ($\sim600-1050$~nm). GRB\,191016A occurred during Sector~17, in a low ecliptic latitude corresponding to the minimum $\sim$27-day baseline. The photometric precision of the light curves are 0.1-1\% at 15-16th magnitude, the peak brightness of this burst. With the start of the TESS extended mission on July 5, 2020, TESS has increased its monitoring cadence from 30~minutes to 10~minutes.

TESS does not release reduced light curves for all regions of the sky. Instead, the mission releases the full-frame images (FFIs) for each cadence in a sector, and requires observers to perform their own photometric extraction, background subtraction and systematic error corrections. The optimal way to do this can differ based on the position on the detector, crowdedness of the source region, the magnitude of the source, and nature of the variability being studied \citep[see, for example, the challenges in adapting the \emph{Kepler} light curves for use on extragalactic targets; ][]{Smith2019}. 

\begin{figure*}
\centering
\begin{tabular}{c}
 \includegraphics[width=\textwidth]{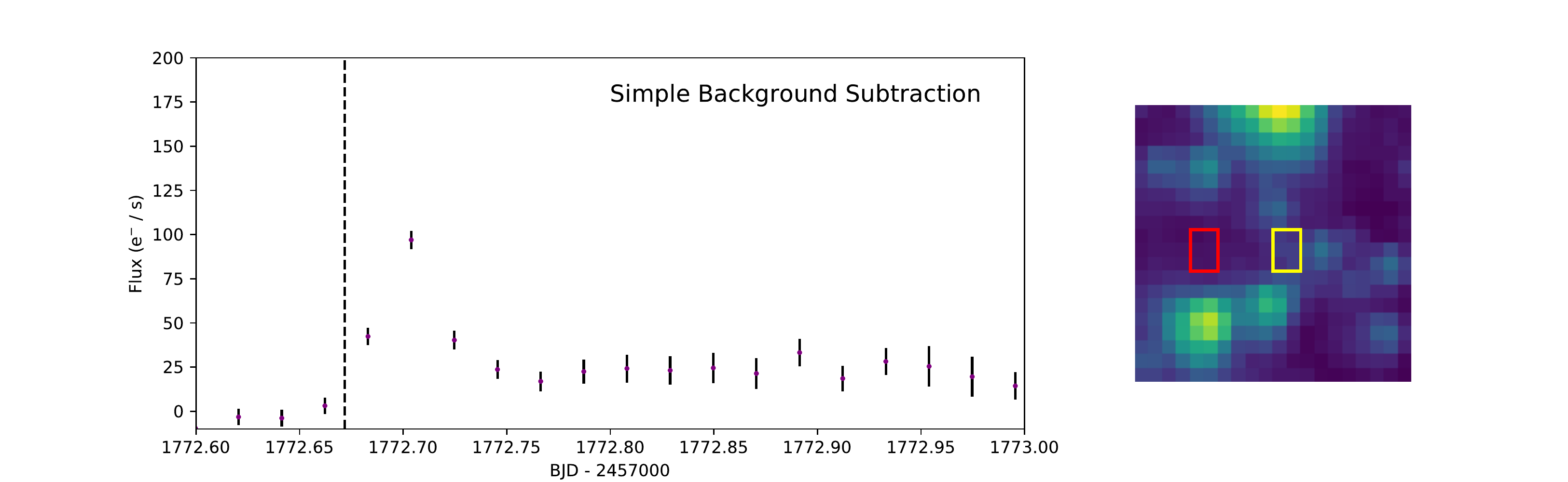} \\
 \includegraphics[width=\textwidth]{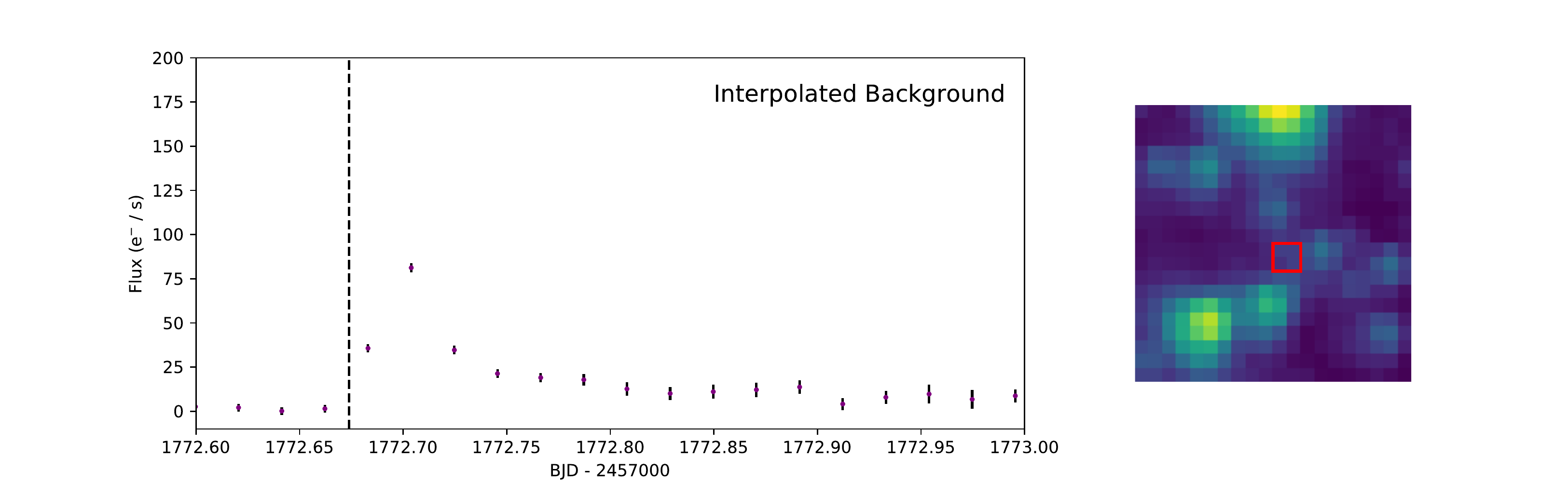}\\
 \includegraphics[width=\textwidth]{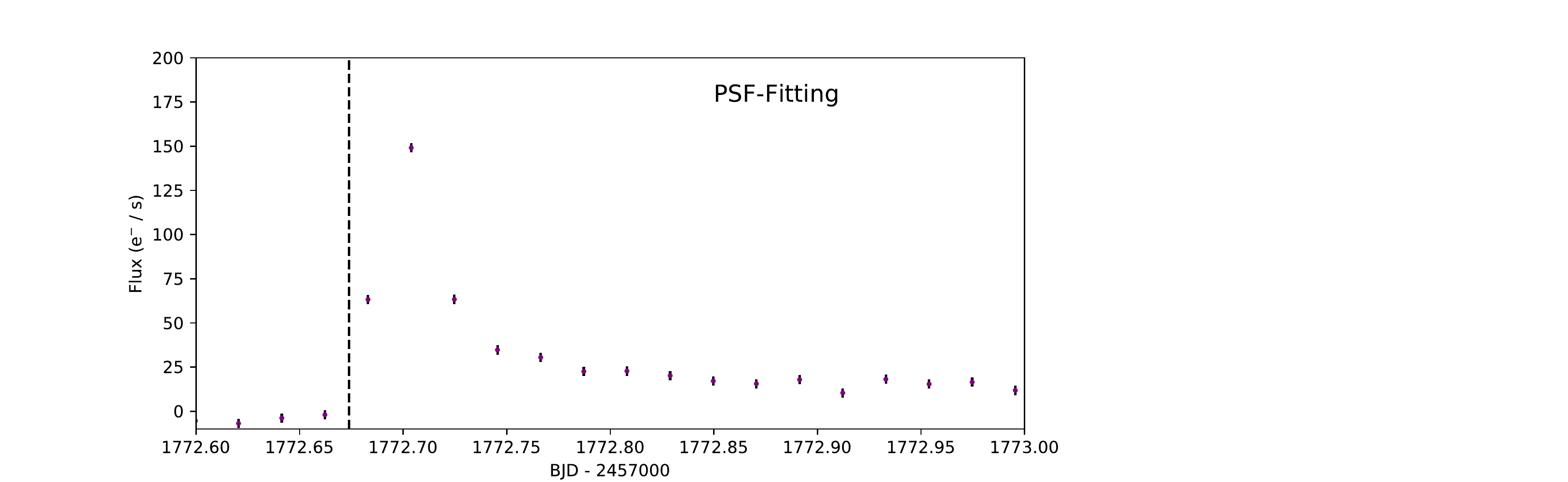}\\
\end{tabular}
\caption{Light curves (left) and extraction and background apertures (right) of GRB~191016A. The first row shows the light curve reduced with simple background subtraction as derived in Section~\ref{sec:smith_lc}, the second row shows the light curve derived using the interpolated background estimation method described in Section~\ref{sec:harper_lc}, and the third row shows the light curve derived  using the PSF-fitting method described in Section~\ref{sec:daylan_lc} (because the PSF is fitted at each cadence, no stable aperture exists). All light curves are normalized by subtracting the median of the first 100 cadences of the sector (not plotted), well before the \emph{Swift} trigger, which is denoted by the vertical dashed lines.}
\label{fig:tess_lc}
\end{figure*}

The TESS satellite is optimized to efficiently search for exoplanet transit signals around more than 20,000 stars simultaneously. This mission is not designed for deep, high resolution studies of individual objects. The TESS angular resolution is extremely low, with pixels measuring 21\arcsec~ across, frequently resulting in a situation where no photometric extraction aperture is possible that does not include incidental nearby sources. Extraction of the light curve is thus dependent upon the source properties desired, the sky environment of the source, and the location on the TESS detector. To ensure that any conclusions we draw in our analysis are robust against different methods, we extract the light curve in three different ways, described in this section.

\subsection{Light Curve Extractions: Simple Background Subtraction}
\label{sec:smith_lc}

In this simplest method of systematics mitigation, we begin by requesting a cutout of the sky around the region of interest using the online TESSCut tool\footnote{https://mast.stsci.edu/tesscut/} \citep{Brasseur2019}; this produces a FITS file with photometric images at every cadence during the monitored sector(s), with all of the information provided in the full FFIs (e.g., quality flags and time/flux errors). The bulk of the manipulation of the FITS files is accomplished with the AstroPy library \citep{astropy1,astropy2}. We then choose an extraction aperture, by eye, to maximize the flux from the GRB afterglow while still avoiding nearby sources, as well as a nearby background region devoid of sources with the same pixel size as the extraction aperture. We perform aperture photometry from each cadence that is not flagged for pointing instability by the TESS mission, from both the extraction and background apertures. Finally, the background light curve is subtracted from the source light curve. The chosen source and background apertures and the resulting light curve are shown in the first row of Figure~\ref{fig:tess_lc}. Note that Figure~\ref{fig:tess_lc} shows the light curves in TESS instrumental units, as is usual for TESS photometric studies. There is currently no accepted method of converting TESS counts to a magnitude system. The light curve is presented again, in a context comparative to other afterglows, in Section~\ref{sec:latepeak}, in more traditional units for GRBs, with an attempt at a magnitude conversion.

\subsection{Light Curve Extraction: Interpolated Background}
\label{sec:harper_lc}
For this method, we interpolate the TESS background from sky pixels. We also utilise TESSCut \citep{Brasseur2019} in conjunction with the Kepler/TESS reduction packages from Lightkurve \citep{lightkurve} for this method, to get a $50\times50$~pixel TESS image of the region surrounding the GRB. To identify which pixels are background pixels we simulate a $50\times50$~pixel TESS image, centered on the GRB, using the Gaia source catalogue \citep{GaiaDR2}. The Gaia sources are mapped onto the simulated image with the TESS WCS. We convert Gaia source magnitudes to TESS magnitudes via $m_{TESS} = m_{Gaia} - 0.5$ \citep{Stassun2018}, and the subsequent TESS magnitudes to counts, with a zeropoint of $20.44$. We then convolve the Gaia sources with a model TESS PSF, using methodology based on the \emph{DAVE} pipeline \citep{Kostov2019}. Finally, all pixels of the simulated image with counts less than a limit are selected as background pixels. We interpolate the background signal from the background pixels to all pixels in the image. This method of simulating the image allows for a clear determination of which pixels are dominated by the background. 

Following background subtraction simple aperture photometry is performed with a $2\times 2$~pixel aperture. The resulting light curve is shown in in the middle panel of Figure~\ref{fig:tess_lc}.

\subsection{Light Curve Extraction: PSF-Fitting}
\label{sec:daylan_lc}

Modeling the target flux via PSF yields an alternative method to extract the flux of the target in each time bin. Towards this purpose, we perform PSF photometry of the target given the TESS full frame image data. We employ the Pixel Response Function (PRF) model constructed by the TESS mission. This takes into account the optical PSF as well as the pointing jitter of the detector. We then use $13\times13$ pixel PRF models that are upsampled by a factor 9. We interpolate the model up to the third order in subpixel shifts and evaluate the model images of the target and nearby known point sources in the TESS Input Catalog \citep[TIC, ][]{Stassun2018}. Finally, using these point source models as templates, we perform linear regression to infer the fluxes of all sources in each time bin. Inclusion of the templates of the nearby sources enables marginalization over the neighbors and reduces potential flux contamination from nearby sources.

\begin{figure*}[t]
\centering
\begin{tabular}{ccc}
 \includegraphics[width=0.3\textwidth]{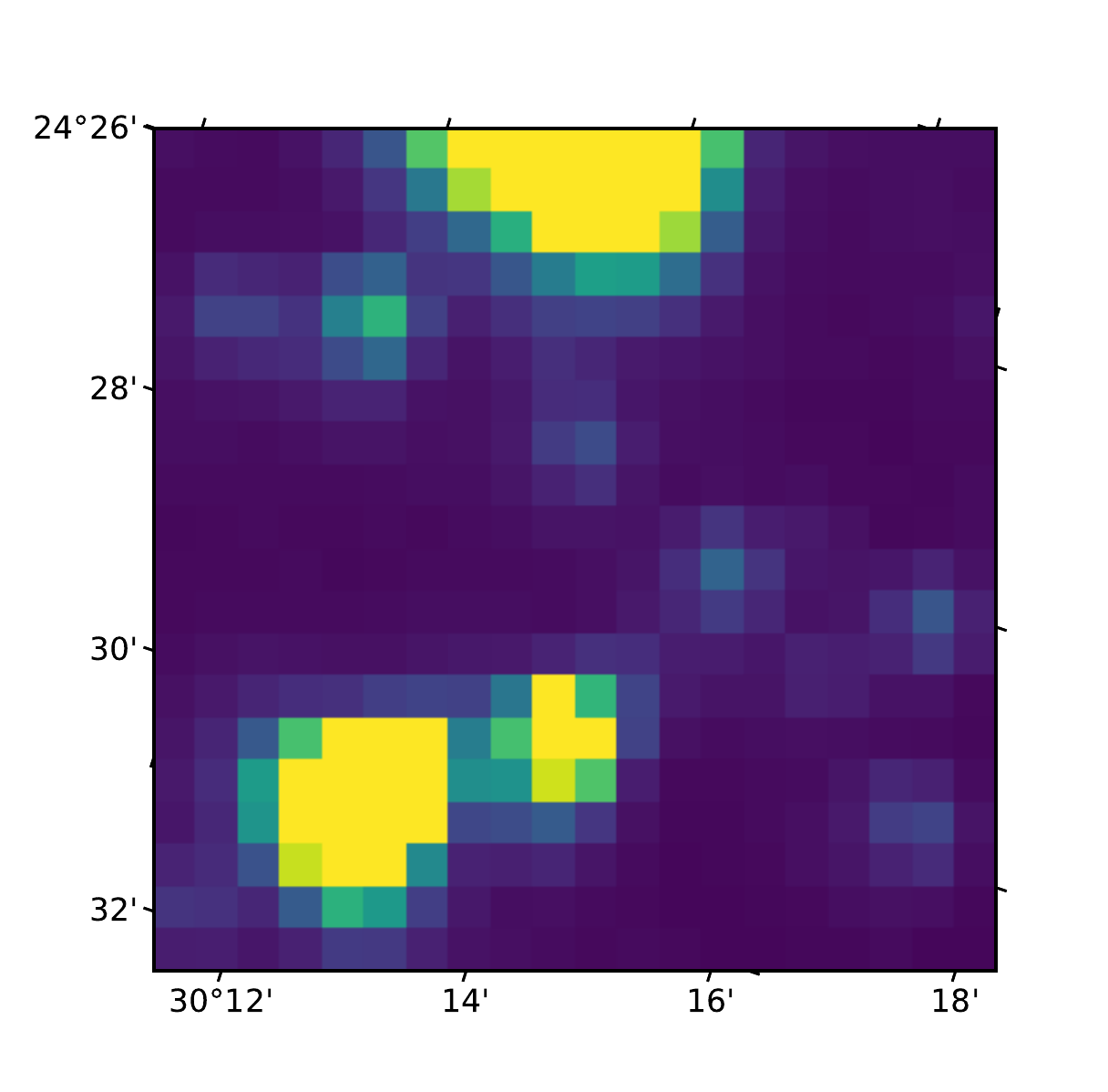} & 
 \includegraphics[width=0.3\textwidth]{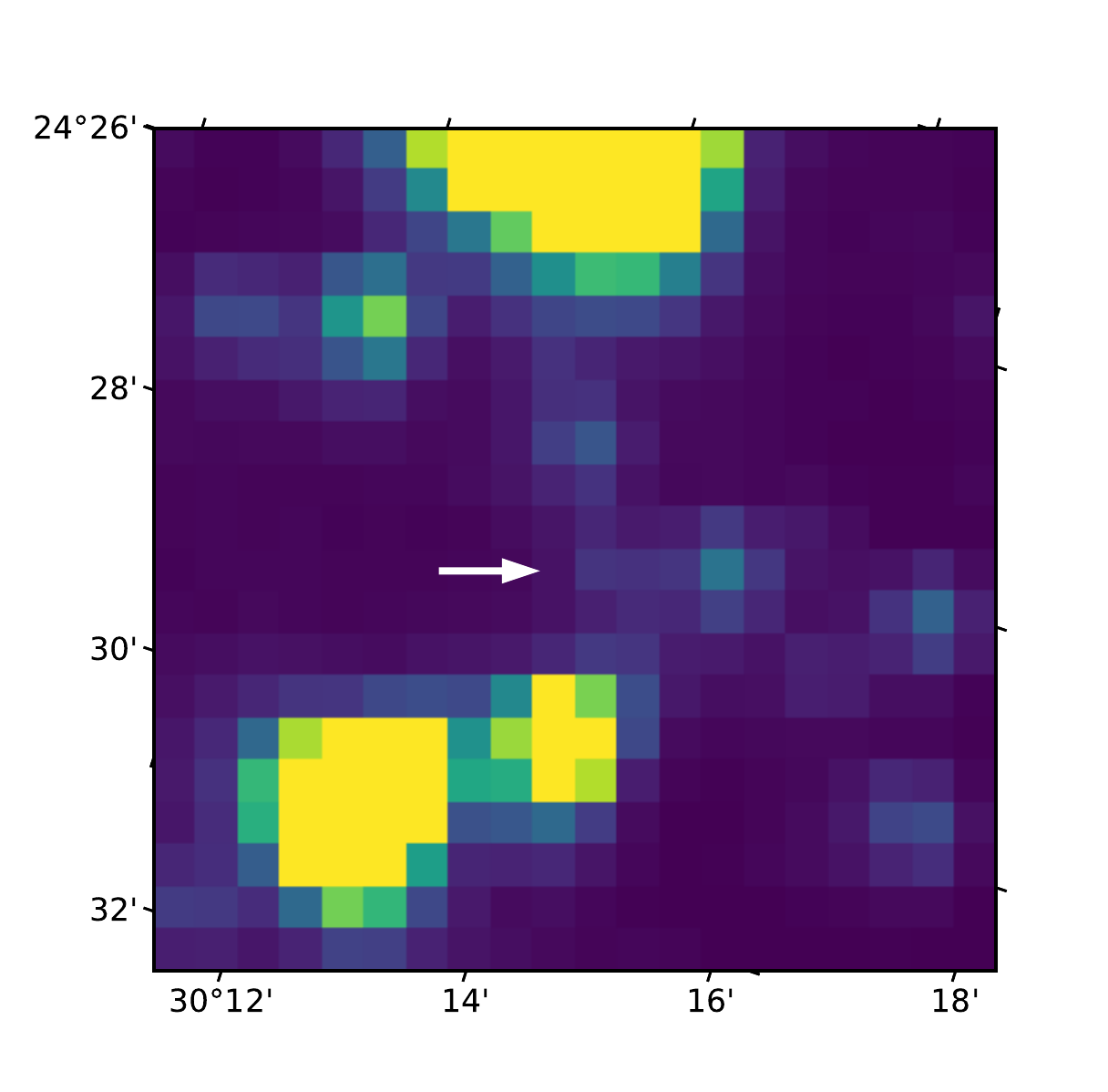} &  \includegraphics[width=0.21\textwidth,trim=0 -1.4cm 0 0]{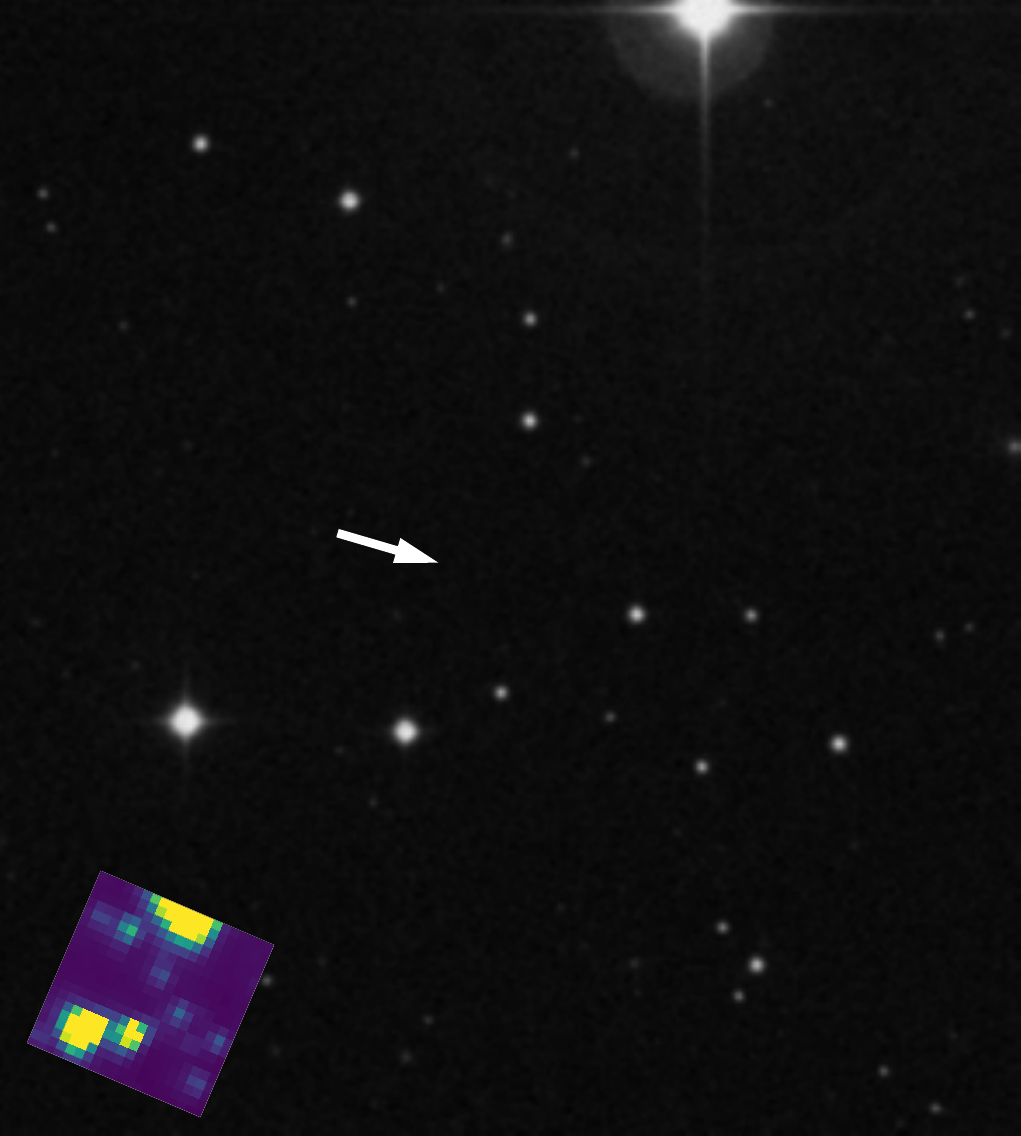}\\
\end{tabular}
\caption{TESS full-frame image in the cadence just before the BAT trigger (left) and at the  peak flux of the burst (center). The emergence of the afterglow is apparent in the center of the image, indicated by the white arrow. Contrast has been increased versus the right panel of Figure~\ref{fig:tess_lc} in order to increase visibility. The right panel shows the same region of the sky, with a slightly different orientation, in the Digitized Sky Survey (DSS); a small inset of the TESS image is provided in the bottom left corner to demonstrate the change in orientation.}
\label{fig:tess_ffis}
\end{figure*}

Each light curve and extraction aperture is plotted in Figure~\ref{fig:tess_lc}.  The light curve is clearly rising in the first data point following the BAT trigger, peaks at the subsequent point, and then declines. The afterglow falls below the TESS limiting magnitude as the light curve flattens out; note that the post-burst baseline may be higher than the pre-burst baseline due to an increase in the lunar scattered light. After the TESS limit is reached, ground-based follow-up is needed to explore the afterglow evolution at fainter magnitudes than $V\sim18$~mag. In Figure~\ref{fig:tess_ffis}, we show the sky image from TESS in the cadence before the BAT trigger and at the TESS peak cadence, with contrast enhanced for easier visibility. The extraction aperture shown in Figure~\ref{fig:tess_lc} encompasses the new afterglow source PSF as completely as possible without including the nearby targets.

\section{High Energy Emission}
\label{sec:fermi}

\begin{figure}
\centering
 \includegraphics[width=0.5\textwidth]{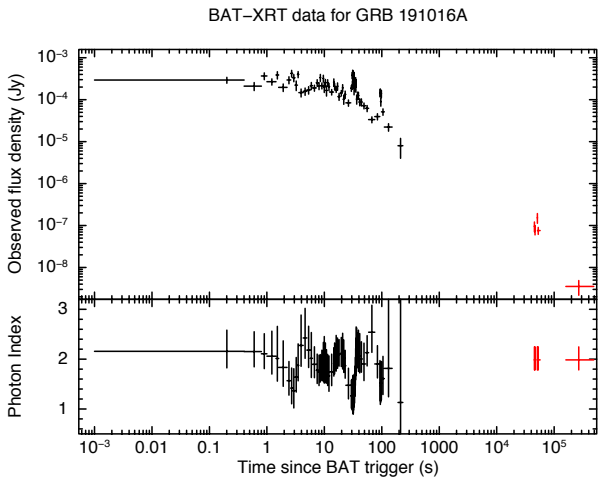}

\caption{The \emph{Swift}-BAT (black) and XRT (red) light curve for GRB~191016A, binned for S/N = 5. The evolution of the photon index is shown in the lower panel. This figure was obtained from the \emph{Swift} Burst Analyser \citep{Evans2010}.}
\label{fig:batlc}
\end{figure}

\begin{figure}
\centering
 \includegraphics[width=0.5\textwidth]{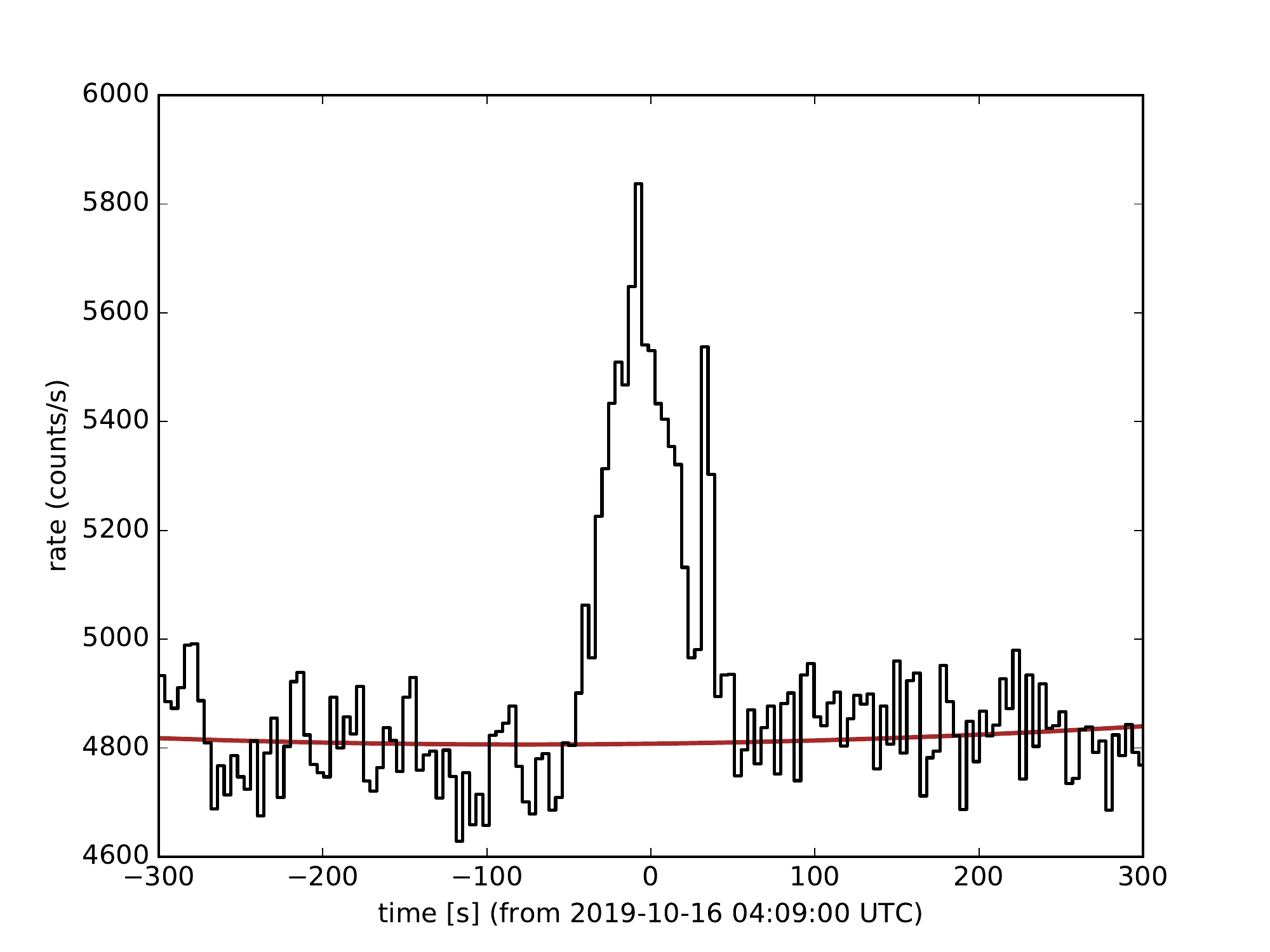}

\caption{The summed \emph{Fermi}-GBM lightcurve from NaI detectors 7, 9, a and b. The energy range is $50-300$~keV and the temporal resolution is 4~s. The brown curve is the fitted polynomial background.}
\label{fig:gbm}
\end{figure}

In this section we discuss the gamma ray emission from GRB~191016A. Although undetected by \emph{Fermi}-LAT, the burst was detected by both \emph{Swift}-BAT (the trigger) and \emph{Fermi}-GBM (no trigger). The BAT light curve is shown in Figure~\ref{fig:batlc}, obtained from the \emph{Swift} Burst Analyser\footnote{\url{https://www.swift.ac.uk/burst\_analyser/00929744/}} \citep{Evans2010} with a S/N = 5 binning. The GBM light curve is shown in Figure~\ref{fig:gbm}. 

High-energy emission from GRBs has been observed by \emph{Fermi}-LAT up to $>$10~ks post-trigger \citep{2019ApJ...878...52A}. We preform an unbinned likelihood analysis in the time window between 3.9~ks and 5.4~ks, when the GRB was in the LAT FoV, in the energy range of $0.1-1$~GeV.
The GRB spectrum is described with a simple power law, and the contribution of the galactic and extra-galactic diffuse emissions are added to the model. 
We select events within a Region Of Interest (ROI) of 12 degrees from the GRB localization.
The time interval is selected by requiring that the entire ROI is visible, with a zenith angle $<$100 degrees. No significant emission at high energy is detected. We compute a flux upper limit of 1.6$\times10^{-6}$ photon cm$^{-2}$ s$^{-1}$, corresponding to an energy flux upper limit of 1.2$\times10^{-9}$ erg cm$^{-2}$ s$^{-1}$ assuming a photon index of $\Gamma_\mathrm{ph}=2$, typical of LAT GRBs.

Even though GRB~191016A was in its field of view, Fermi-GBM was not triggered. 
The GBM {\tt targeted search} \citep{Goldstein+19targeted} is the most sensitive, coherent search for GRB-like
signals. During the automatic processing of external triggers in the GBM continuous data, the {\tt targeted search} found a significant signal consistent in time, sky location and lightcurve morphology with GRB~191016A. The reason GRB~191016A did not trigger GBM is likely an interplay between different effects: (1) a gradual rise in intensity for this burst (2) the timescale on which GBM calculates the background rates to determine excess (3) variations in the gamma-ray background at the time of GRB 191016A due to the routine slewing of the spacecraft. 

We performed standard spectral analysis using {\tt RMfit}\footnote{\url{https://fermi.gsfc.nasa.gov/ssc/data/analysis/rmfit/}}. We selected data from $T_0-31.7$~s to $T_0+58.5$~s relative to the Swift trigger time. The best fitting spectrum is a power law with an exponential cutoff ($dN/dE\propto E^{-\Gamma_\mathrm{ph}}\exp (-E(2-\Gamma_\mathrm{ph})/E_{\rm peak}$). The photon index $\Gamma_\mathrm{ph}= 1.16 \pm 0.07$ and the energy where the $\nu F_\nu$ spectrum peaks, $E_{\rm peak}=197 \pm22$ keV. The flux in this time interval is $F= (1.04\pm 0.05) \times 10^{-7}$ erg s$^{-1}$ cm$^{-2}$ (10-1000 keV range).
Using the method described in \citet{Bloom2001}, we integrate the spectrum in the canonical 1 keV -10 MeV range and perform the k-correction to obtain the isotropic-equivalent energy in gamma rays. We calculate $E_{\gamma,{\rm iso}} =(2.37\pm 0.12) \times 10^{53}$ erg.


\section{A Late-Peaking Afterglow}
\label{sec:latepeak}

The brightest point in the TESS light curve occurs at 2589.7~s after the BAT trigger. This is significantly later than a typical long GRB in which the optical rise is observed; \citet{Oates2009} find that in their sample of 27 UVOT afterglows, all light curves are decaying by 500~s after trigger in the observer's frame. Such a late peak is not unprecedented, however; peak times of $\sim10^3$ seconds are still within the tail of the distributions reported by \citet{Ghirlanda2018} for 67 afterglows with observed peaks.  

\begin{figure}
\centering
 \includegraphics[width=0.5\textwidth]{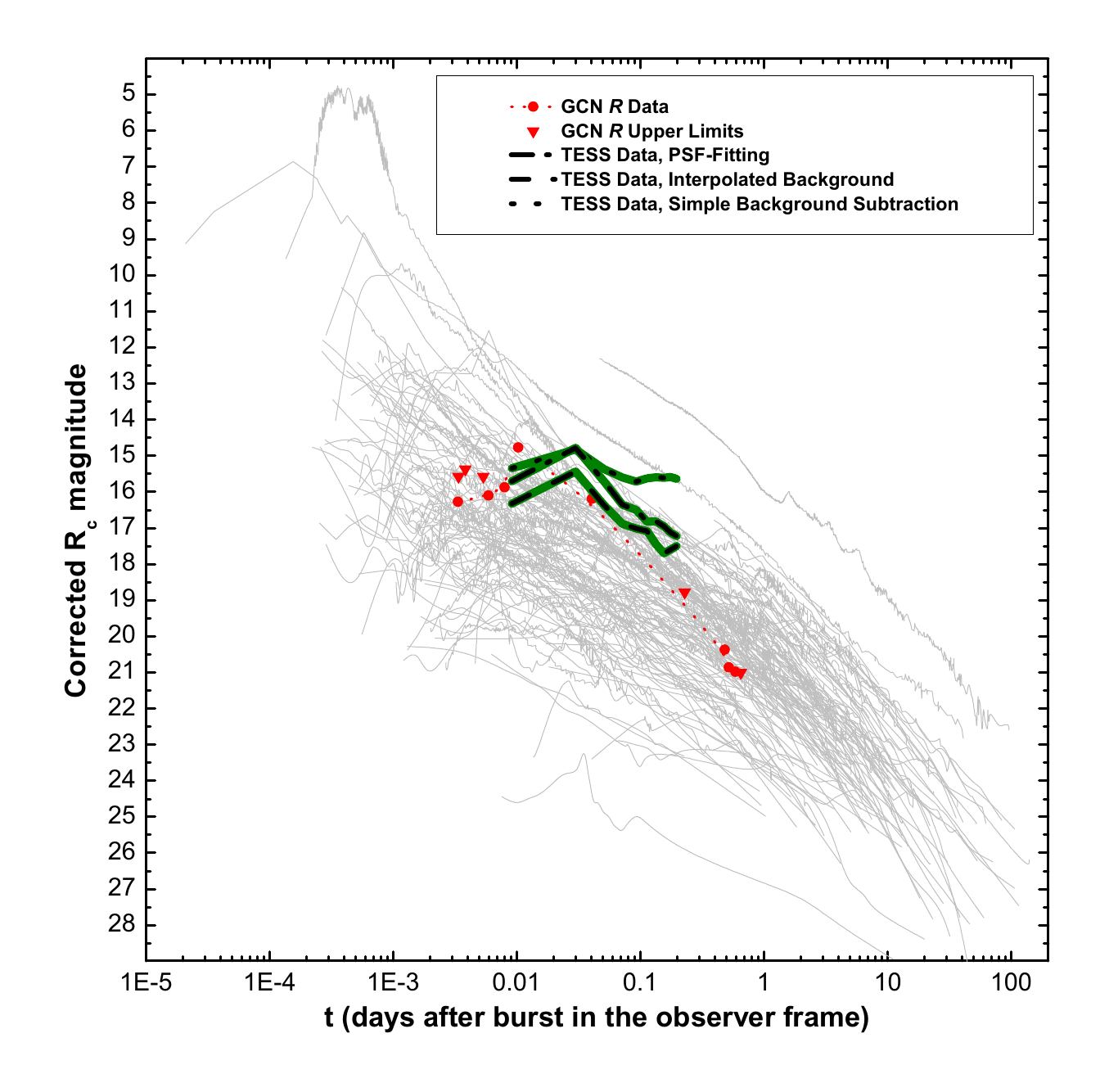}

\caption{Optical afterglows of GRB~191016A from the three TESS approaches (Section~\ref{sec:tess}; c.f. Figure~\ref{fig:tess_lc}), shown in green, compared to the long GRB afterglow sample from \citet{Kann2010}. Also shown is the ground-based $R$-band light curve from the GCN, in red.}
\label{fig:kannplot}
\end{figure}

In Figure~\ref{fig:kannplot}, we show the three TESS afterglow light curves as computed in Section~\ref{sec:tess} and the ground-based GCN light curve, overlaid upon the large sample of optical afterglows of long GRBs from \citet{Kann2010}. Although the TESS light curves are relatively bright, neither their peak magnitudes nor late peaking time is far out of the ordinary. \citet{Kann2010} (their Figure 7) also show some examples of afterglows that peak at significantly later times, including GRB~970508 which peaked after one day.

A very important caveat is necessary when comparing the brightness of the TESS afterglow to others in the plot, however: there is currently no accepted conversion from TESS counts (as shown in Figure~\ref{fig:tess_lc}) and any conventional magnitude system. This is primarily due to the monolithic and unusual TESS bandpass. In order to make a reasonable approximation, we have used the statement in the TESS Instrument Handbook\footnote{https://heasarc.gsfc.nasa.gov/docs/tess/documentation.html} that 15,000~e$^{-}$s$^{-1}$ are generated by the cameras for a star of apparent magnitude $m=10$. We have thereby calculated the ``zero flux" of the TESS bandpass to be $\sim1.5\times10^{8}$~e$^{-}$~s$^{-1}$, via $m = -2.5 \mathrm{log} (F/F_0)$. This number is roughly consistent with what is obtained by comparing the TESS light curves to the ground-based light curves, although the points are not simultaneous. Due to the roughness of this approach, the vertical normalization of the light curves on Figure~\ref{fig:kannplot} should be assumed to have large uncertainties.

Each of the TESS light curves yields a different temporal decay index $\alpha$, where $F\propto t^{\alpha}$: $\alpha = -0.3$ for Simple Background Subtraction, $\alpha = -1.1$ for PSF Fitting, and $\alpha = -1.0$ for the Interpolated Background method. Since the Simple Background Subtraction method is most likely to suffer from contamination at later times, due to a large extraction aperture and no background modeling, its shallower index is not surprising. The decay index from the ground-based data is -1.4. These are all within the typical values for optical afterglows, as can be seen in Figure~\ref{fig:kannplot}. In comparison to the optical afterglow decay indices of a sample of 139 long GRBs by \citet{DelVecchio2016}, the value for simple background subtraction, -0.3, is in the very shallow end, but the other values are among the most commonly found. 

Some individual objects have been observed with late-peaking afterglows, typically following a long plateau phase in the optical and X-ray light curves. We do not have X-ray data during the optical rise time, since XRT did not slew to the position in time (Section~\ref{sec:grb}), so we do not know whether the X-ray flux was consistent with this plateau behavior. There is only a single TESS data point between the trigger time and the apparent peak, which occurs 786~s after the trigger; already quite late compared to the \citet{Oates2009} sample. It is possible that this point occurs during an optical plateau, after which the flux rises rapidly to the peak data point before steeply decaying; the ground-based light curves do not indicate such a flattening, but do not definitively rule it out, either. If this is the case, it resembles the UVOT light curve of GRB~100418A, which peaked very late at approximately 50ks post-trigger, and was determined to most likely arise from continuous injection of energy into the forward shock \citep{Marshall2011}, as was the optical afterglow of GRB~060729 \citep{Grupe2007}. Indeed, a recent re-analysis of GRB~100418A by \citet{deUgarte2018} showed that the afterglow re-brightened rapidly during the first day after the trigger, beginning 2.4~h after the burst. Other late-peaking afterglows have been interpreted as due to an off-axis viewing angle, where the peak occurs once the beam has widened sufficiently to include the line of sight, as in GRB~080710 \citep[$t_\mathrm{peak} \sim 2\times10^3$~s, ][]{Kruhler2009} and GRB~081028  \citep[$t_\mathrm{peak}\sim3\times10^4$~s, ][]{Margutti2010}. The lack of a detection of GRB~191016A by \emph{Fermi}-LAT may also support the off-axis interpretation for this burst; however, if the photometric redshift calculated below (Section~\ref{sec:redshift}), $z_\mathrm{phot} = 3.29$, is correct, a nondetection is perhaps to be expected, since only one GRB at a higher redshift has been detected by \emph{Fermi}-LAT \citep[GRB 080916C at $z=4.35$, ][]{fermi2009}.

Late-peaking or complex afterglows can also be the result of a number of other physical scenarios, including reverse or forward shocks due to interaction with the ISM or progenitor winds \citep{Sari1999,Kobayashi2004}, the peak frequency of the synchrotron emission moving through the observing band \citep{Sari1998}, or destruction of surrounding dust by radiation as the burst proceeds \citep{Fruchter2001}; see \citet{Oates2009} for a nice summary of these effects in greater detail.

\section{Estimating the Redshift}
\label{sec:redshift}

The only constraint placed on the redshift by the Gamma-ray Coordinates Network (GCN) circulars was $z < 4$, established by the RATIR team \citep{Watson2019b}. 

\begin{figure}
\centering
 \includegraphics[width=0.5\textwidth]{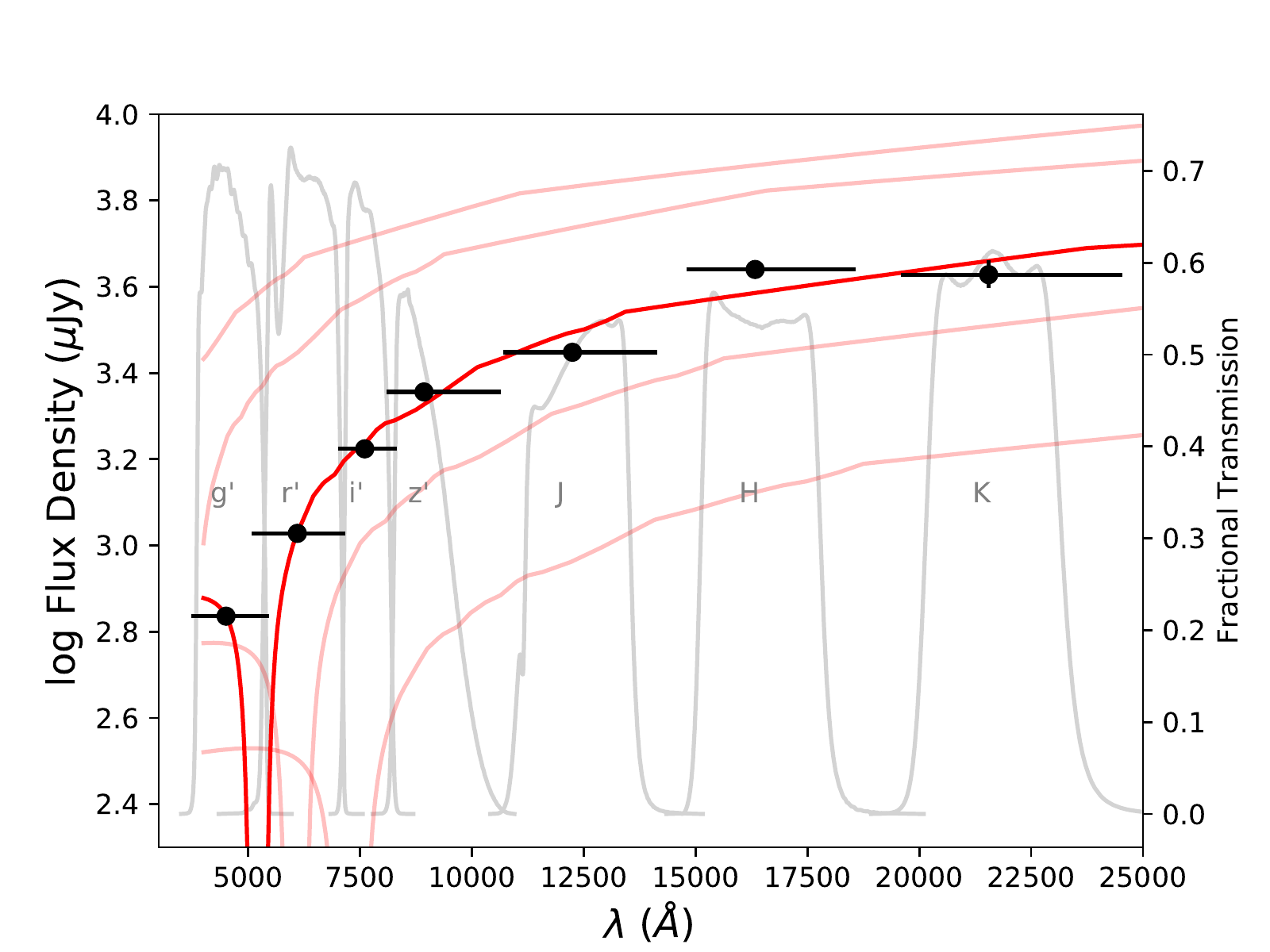}

\caption{Spectral energy distribution from the GROND instrument at La~Silla Observatory, the filter curves of which are shown. The best-fitting model corresponds to $z_\mathrm{phot}=3.29$, and is shown in red. Other parameters are discussed in Section~\ref{sec:redshift}. Four other models are shown in pink, with identical parameters to the best-fitting model, but with redshifts 1, 2, 4, and 5 from top-left to bottom-right.}
\label{fig:sed}
\end{figure}

In order to determine a more precise redshift, we make use of the GCN-reported GROND photometry \citep{Schady2019}, plotted in Figure~\ref{fig:sed}. Since we lack UVOT photometry, we do not have the fullest SED possible to constrain the redshift. However, with a simple model, we can arrive at a much more precise estimate than from the $g-r$ color alone.

We follow the prescription laid out in \citet{Kruhler2011}, which we describe briefly here, along with our simplifications. We begin by assuming that the intrinsic shape of the SED is a power law: $F_\nu (\lambda) = F_0 \lambda^{\beta_0}$, where $F_0$ is a normalization constant. This intrinsic power law is then modified by extinction in the host galaxy, and by any neutral hydrogen in the intergalactic medium along the line of sight. Many GRB afterglows exhibit damped Lyman-$\alpha$ (DLA) absorption associated with the host galaxy, which must also be accounted for. The effect of extinction from within the Milky Way is negligible: N$_\mathrm{HI,MW} = 7.57\times10^{20}$~cm$^{-2}$ \citep{BenBekhti2016}. After these effects are accounted for, the observed spectrum can be modelled as

\begin{equation}
    F_\nu (\lambda) = D_\alpha(z) F_0 \lambda^{\beta_0} \mathrm{exp}[-\tau_\mathrm{dust}(z,A_V) - \tau_\mathrm{DLA}(z,N_H)].
\end{equation}

\noindent This model includes four free parameters: the redshift $z$, the intrinsic power law index $\beta_0$, the host galaxy extinction $A_V$, and the host column density of neutral hydrogen, $N_H$. 

$D_\alpha(z)$ is the wavelength-averaged attenuation due to line blanketing from intergalactic neutral hydrogen. This is a monotonically-increasing function with redshift, which we model as in \citet{Madau1995} (see their Figure~2). The first optical depth term, $\tau_\mathrm{dust}$, accounts for the host galaxy's own dust reddening. As described by \citet{Kruhler2011}, most bright GRB afterglows are well-modelled by a local reddening law ($A_\lambda / A_V$ as a function of wavelength) based on the Small Magellanic Cloud, as opposed to models based on the Large Magellanic Cloud and the Milky Way; we nonetheless attempt the modelling using each of the three extinction laws, reproduced from \citet{Pei1992}; the lowest $\chi^2$ values are found using the SMC version. With the reddening law in hand, we then follow \citet{Li2018}: $\tau_\mathrm{dust} = (1/1.086) A_V \eta(\nu)$, where $\eta(\nu) = A_\lambda / A_V$. 

The second optical depth term accounts for neutral hydrogen within the host galaxy ($\tau_\mathrm{DLA}$). This term is calculated following \citet{Totani2006}: $\tau_\mathrm{DLA}(\lambda_\mathrm{obs}) = N_H \sigma_\alpha[\nu_\mathrm{obs} ( 1+z)]$, where $\nu_\mathrm{obs} = c/\lambda_\mathrm{obs}$, and $\sigma_\alpha$ is the exact formula for the frequency dependence of the Ly$\alpha$ cross-section \citep[e.g., ][]{Madau2000}. 

We fit Equation~1 to the SED, using the observed fluxes and mean wavelengths in each filter, with the definitions above using a Nelder-Mead minimization  \citep{Nelder65}, allowing $z$, $\beta$, log~$N_H$, and $A_V$ to vary freely. We find that the best-fit model is achieved with $z_\mathrm{phot} = 3.29\pm{0.40}$, $\beta_0 = 0.16\pm{0.02}$, log~$N_H = 23.0\pm{0.97}$, and log~$A_V = -0.45\pm{0.12}$, which achieves a reduced $\chi^2 = 4.1$. The uncertainties in the fit parameters were estimated by using the results of the Nelder-Mead fit as the basis of a new least-squares fit, which generates a covariance matrix. The parameter errors are then the square root of the product of the diagonal of this matrix and the reduced chi-squared value of the least-squares fit.

These parameters are within observed distributions for GRB host galaxies as reported by \citet{Li2018}. The value for $N_H$ is quite high, but with a large uncertainty. Such values have been observed in other bursts, such as GRB~080607 ($z = 3.04$), and can be attributed either to molecular clouds along the line-of-sight in the host, or to the large-grained nature of dust at $z\sim3$; see \citet{Corre2018} for a discussion.

The SED, GROND filters, and best-fitting model are shown in Figure~\ref{fig:sed}, along with a few other models for comparison.

\section{Bulk Lorentz Factor}
\label{sec:lorentz}

If indeed one of the above scenarios causes the late-peaking afterglow and the the burst is on-axis, with a peak time
corresponding to the start of the afterglow emission,
one can derive the bulk Lorentz factor of the outflowing material using the equation and normalizations given by \citet{Molinari2007}:

\begin{equation}
    \Gamma_\mathrm{BL}(t_\mathrm{peak}) =  \left[
    \frac{3E_\gamma(1+z)^3}{32\pi n m_\mathrm{p}c^5\eta t_\mathrm{peak}^3}  \right] ^{1/8} \approx 160 \left[ \frac{E_{\gamma,53}(1+z)^3}{\eta_{0.2}n_0t_\mathrm{peak,2}^3} \right]^{1/8},
\label{eq:gamma}
\end{equation}

\noindent where $E_{\gamma,53}$ is the isoptropic-equivalent energy released in gamma rays normalized to $10^{53}$~erg, $n_0$ is the particle density of the surrounding medium in cm$^{-3}$, the normalized radiative efficiency $\eta_{0.2}$ is defined as $\eta = 0.2 \eta_{0.2}$, and $t_\mathrm{peak,2} = t_\mathrm{peak} / (100\mathrm{s})$.

To calculate $E_\gamma$, the isotropic-equivalent energy released by the burst in gamma rays, we use the relation from \citet{Bloom2001}:

\begin{equation}
E_\mathrm{iso} = S \frac{4\pi D_l^2}{1+z} k,
\end{equation}

\noindent where $S$ is the fluence, $D_l$ is the luminosity distance and $k$ is the cosmological $k$-correction factor, which is approximately 2.5 at the best-fitting model redshift of $z_\mathrm{phot}=3.29$. Using this and the fluence reported by the BAT detection ($1.12\times10^{-5}$~erg~cm$^{-2}$; see Section~\ref{sec:grb}), we find that $E_\mathrm{iso} = 6.61\times10^{53}$~erg. This is consistent with the most populous region of the observed range of $E_\mathrm{iso}$ found in a sample of 92 long GRBs by \citet{Ghirlanda2009}. It is slightly higher than the value of $E_\mathrm{iso}$ found in Section~\ref{sec:fermi}; this inconsistency likely results from the longer observation time and smaller energy window for \emph{Swift} ($15-350$~keV), and the fact that $E_\mathrm{peak}$ is not constrained by \emph{Swift} observations. Extrapolating  a power law from \emph{Swift's} energy range to the 1~keV-10~MeV with no cutoff naturally results in a larger $E_\mathrm{iso}$. We therefore have two values of $E_\mathrm{iso}$ that can be used in Equation~2 for Lorentz factor calculation. 

In addition to $E_\mathrm{iso}$, Equation~2 requires an estimate of $t_\mathrm{peak}$ to determine the bulk Lorentz factor $\Gamma_\mathrm{BL}$. We may either take the TESS peak at face-value and declare it the true peak time, or use extrapolation to determine the earliest possible peak time. The time between the BAT trigger and the brightest data point in the TESS light curve is 2589.7~seconds. Of course, the brightest TESS point may not represent the actual peak time, especially since the TESS cadence is a relatively coarse 30 minutes. The light curve is clearly still rising at the previous TESS data point, which occurs 786~s after the BAT trigger, providing a lower limit on $t_\mathrm{peak}$. 

Based on the fact that the observed peak is relatively late, we may instead assume that the true peak occurred somewhere between the observed peak and the previous still-rising data point. \citet{Oates2009} calculated the best-fitting power law index to their sample of optical GRB afterglow light curves within the first 500 seconds, $\alpha_{<500}$. The steepest rising light curve in that sample had $\alpha_{<500} = 0.73\pm{0.14}$. If we use this value and extrapolate from the rising data point at $t = 786$s, and then see where that rising power law intersects with the declining power law between the observed peak and the following data point, the inferred peak time occurs at $t=1316$s. Our two estimates for $t_\mathrm{peak}$ are therefore 2590s and 1316s.

If we take the observed peak cadence at $t=2590$s to be the true peak and use the value of $E_\mathrm{iso}$ from Equation~3, the result is $\Gamma_\mathrm{BL} = 103$. This is consistent with the correlation between $\Gamma_\mathrm{BL}$ and $E_\mathrm{iso}$ for long GRBs from \citet{Ghirlanda2018} for a homogeneous ISM. Using the value of $E_\mathrm{iso}$ from Section~\ref{sec:fermi}, we obtain $\Gamma_\mathrm{BL} = 90$.

If we instead take the extrapolated peak time, $t=1316s$, and use the value of $E_\mathrm{iso}$ from Equation~3, we obtain a Lorentz factor of $\Gamma_\mathrm{BL} = 133$. With $E_\mathrm{iso}$ from Section~\ref{sec:fermi}, we obtain $\Gamma_\mathrm{BL} = 117$.

These values of $\Gamma_\mathrm{BL}$ are consistent with the cumulative distribution of bulk Lorentz factors in afterglows with an observed $t_\mathrm{peak}$ for models that assume a homogeneous circumburst medium \citep{Ghirlanda2018}, except for  $\Gamma_\mathrm{BL} = 90$, which is low compared to that distribution.

\section{TESS and Gamma Ray Bursts}
\label{sec:discussion}

In this section, we discuss how the TESS bandpass, sensitivity, sampling pattern and cadence will affect detection rates of GRBs, and to what extent a detection like GRB~191016A can be expected in the future.

The chief advantage TESS offers in studying optical afterglows of GRBs is its continuous coverage independent of a trigger, potentially capturing GRB afterglows serendipitously, with a much higher sampling cadence than other timing surveys like the Zwicky Transient Facility (ZTF) or the Rubin Observatory's Legacy Survey of Space and Time (LSST). This is advantageous in the instance seen here, when observing constraints prevented a rapid slew by \emph{Swift}, providing supplemental photometry to ground-based observations. It would be especially helpful in the case that a GRB does not trigger \emph{Swift}-BAT, providing potentially the only optical follow-up in such cases; for example, for bursts which are only detected by \emph{Fermi}-GBM. 

A major limitation of TESS for GRB follow-up is its much brighter limiting magnitude compared to most ground-based telescopes. The TESS bandpass, as discussed in Section~\ref{sec:tess}, is  red-white monolithic, spanning 600-1100~nm. As such, it does not quite correspond to the traditional ``white" filters on ground based telescopes. TESS uses a self-defined quantity, the ``TESS magnitude," to determine its sensitivity limits, as can be seen in the TESS Instrument Handbook\footnote{https://heasarc.gsfc.nasa.gov/docs/tess/documentation.html}. The photometric precision per 30-minute integration of the TESS cameras falls to $\sim10$\% at apparent magnitudes of $\sim18$~mag \citep{Ricker2015}, and this is only true if a source is isolated and on a well-behaved portion of the CCD uncontaminated by scattered light. 

\begin{figure}
\centering
 \includegraphics[width=0.5\textwidth]{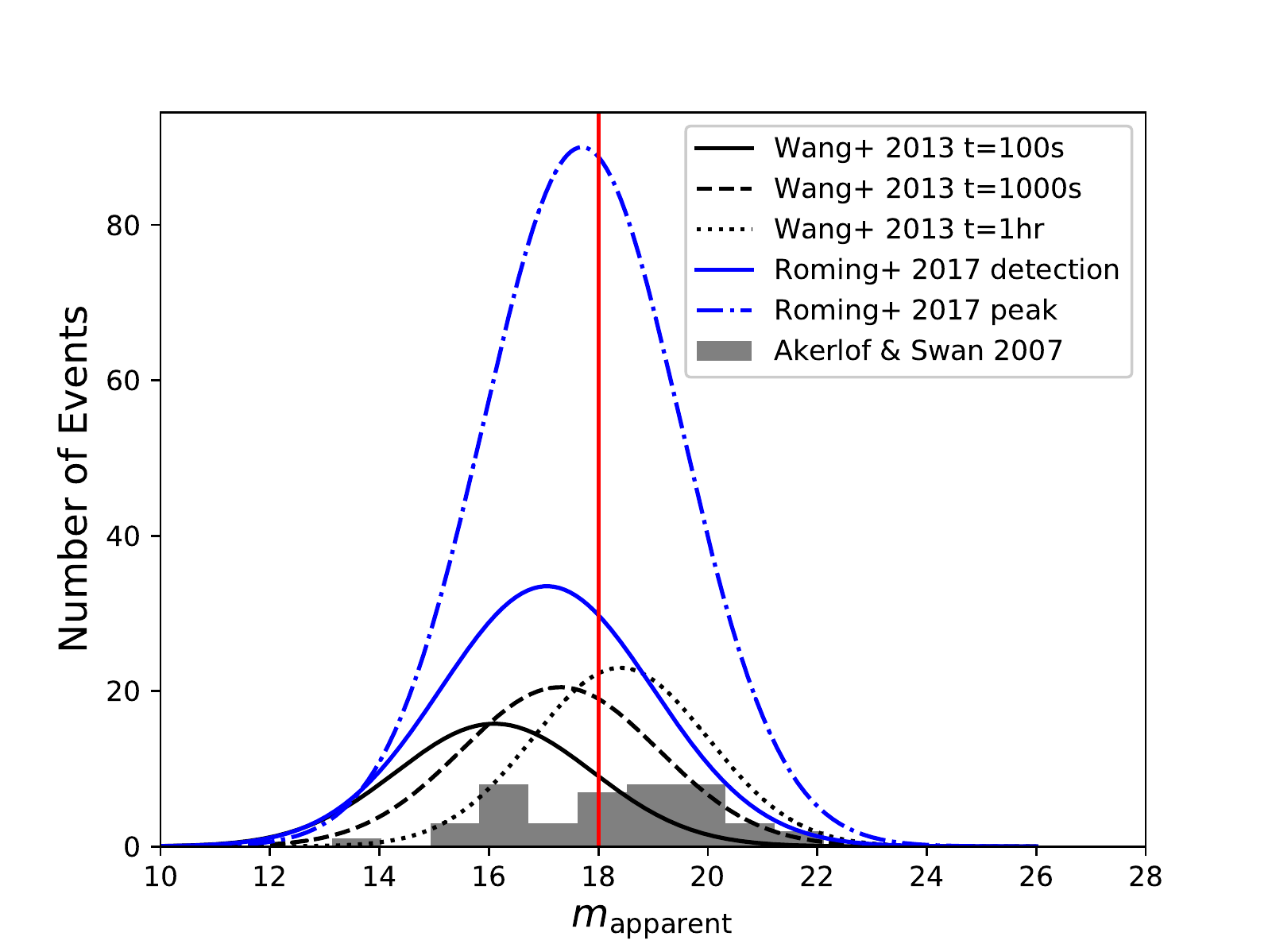}

\caption{Histogram of apparent magnitudes, mostly in the $R$ band but all in bands within the monolithic TESS filter. Blue curves correspond to the distributions reported by \citet{Roming2017}; black curves correspond to the distributions reported by \citet{Wang2013}; the grey histogram represents the distribution from \citet{Akerlof2007}. The TESS magnitude at which photometric precision falls below 10\% is shown by the red vertical line.}
\label{fig:maghist}
\end{figure}

With full frame images spaced every ten minutes, as will be the case starting in TESS Cycle~3, the worst-case scenario is that the burst occurs quasi-simultaneously with a TESS cadence. In this case, the next cadence will occur 10 minutes, or 600 seconds, later. This is after all of the light curves in the \citet{Oates2009} have passed their peak and begun to decay. The distribution of afterglow apparent magnitudes depends on the time after the burst and the morphological type of the light curve. \citet{Akerlof2007} found that the distribution of afterglow apparent magnitudes of \emph{Swift}-detected GRBs peaked at $R \sim 19.5$~mag. \citet{Wang2013} differentiated the observed afterglow population by post-burst time and light curve morphology, finding apparent magnitude distributions peaking at $R=16.1$, 17.3, and 18.4~mag for 100~s, 1000~s, and 3600~s (1~hr) after the burst. \citet{Roming2017} report the UVOT magnitudes of the first-observed data point and the afterglow peak as 17.06 and 17.7~mag, respectively. We reproduce these populations in Figure~\ref{fig:maghist}, and overplot the TESS 10\% photometry limit.

Based on these distributions, approximately 60\% of bursts have afterglows that are above the TESS limiting magnitude for at least one cadence. This is a rough estimate, since the TESS bandpass does not correspond exactly to any of the bandpasses used in the works above. While it would be preferable to know how many bursts will have at least two cadences detected by TESS, precise calculations on this point are not useful, since 18~mag is not a hard limiting magnitude, and under good conditions, fainter objects can be detected by TESS and under poor conditions the limit is brighter.

In the case of GRB~191016A, the peak apparent magnitude is $R=15.1$~mag, and we are able to detect one rising cadence, the peak cadence, and possibly 2 decaying cadences.

The most recent full catalog of UVOT GRB afterglows \citep{Roming2017} contains 626 total bursts detected by multiple gamma ray missions over six years, of which 538 (86\%) were followed up with UVOT. The majority of the remainder were too close to the sun or moon for follow-up, as is the case with GRB~191016A, while a few occurred during instrument downtime. Of the 538 bursts followed up by UVOT, 333 were detected in at least one of its filters. As can be seen in Figure~\ref{fig:maghist}, approximately 60\% of these are brighter than 18th magnitude. Note that although these filters are ultraviolet through $v$, typical burst SEDs are brighter in the red/infrared, and so the probable detection fraction in the redder TESS bandpass is likely higher than 60\%. Additionally, any influence of dust in the Milky Way, the host galaxy, and absorbers in the intergalactic medium are diminished at redder wavelengths. Spread over six years, this means that an occurrence rate of GRBs bright enough for TESS detection is about 33 per year. During a given sector, TESS is surveying approximately 2300 square degrees, or about 4\% of the sky; thus, the likely rate of such a GRB occurring in the TESS field of view is 4\% of 33, or about one per year. In order for TESS to observe it, it must remain above the limiting magnitude for long enough that the TESS 10-minute cadence will capture it. If we assume the very common temporal decay index $\alpha = -1.4$, a burst peaking at $m=16$~mag  will decay below $m=18$~mag in 26s. Bursts peaking at $m=15$ and 14~mag will decay below $m=18$~mag in 138 and 720 seconds respectively; meaning that in the worst case scenario of a burst occurring just before a TESS cadence, only afterglows with relatively bright ($<14$~mag) peaks will be captured. The range of temporal decay indices, however, assures us that approximately half of bursts will have flatter decays than this, remaining brighter for longer. Given the uncertainties in the true limits of TESS detection, along with the possibility of TESS capturing the burst near to its peak and the positive effects of the redder TESS bandpass, a detection rate of one GRB afterglow per year remains a reliable estimate, if perhaps slightly optimistic. It is consistent with the current detection rate.

This discussion is valid only for GRBs powerful enough to trigger \emph{Swift}, since the comparison samples in Figure~\ref{fig:maghist} are based on \emph{Swift} triggers, and with sufficient positional accuracy to localize to a TESS source. It is possible that the TESS-GRB detection rate may be slightly higher, if, for example, orphan afterglows are taken into account.

\section{Conclusion}
\label{sec:conclusion}
We have analyzed the TESS light curve and ground-based photometry of GRB~191016A, a long GRB detected by \emph{Swift}-BAT. The afterglow has a late peak that is at least 1000~seconds after the BAT trigger, with a brightest-detected TESS datapoint at 2589.7~s.  Using photometric modelling, we have determined the redshift of the afterglow to be $z_\mathrm{phot} = 3.29\pm{0.40}$. The burst was not immediately observed by the XRT and UVOT due to its proximity to the moon, which is true of about 14\% of \emph{Swift} bursts; the serendipitous ongoing monitoring of TESS therefore provided prompt follow-up within a few minutes that otherwise would have been missed for this burst, supplementing triggered ground-based observations. Simple arithmetic arguments based on archival afterglow samples imply that TESS will likely detect $\sim1$ GRB afterglow per year above its magnitude limit, not accounting for afterglows without \emph{Swift} triggers.

\acknowledgments

Support for KLS was provided by the National Aeronautics and
Space Administration through Einstein Postdoctoral Fellowship
Award Number PF7-180168, issued by the Chandra
X-ray Observatory Center, which is operated by the
Smithsonian Astrophysical Observatory for and on behalf
of the National Aeronautics Space Administration
under contract NAS8-03060. P.V. acknowledges  support from NASA grants 80NSSC19K0595 and NNM11AA01A. This paper includes data collected by the TESS mission. DAK acknowledges support from Spanish National Research Project RTI2018-098104-J-I00 (GRBPhot). Z.R.W. acknowledges support through NASA grant 80NSSC19K1731. Funding for the TESS mission is provided by the NASA Explorer Program. This work has made use of data from the European Space Agency (ESA) mission {\it Gaia} (\url{https://www.cosmos.esa.int/gaia}), processed by the {\it Gaia}
Data Processing and Analysis Consortium (DPAC, \url{https://www.cosmos.esa.int/web/gaia/dpac/consortium}). Funding for the DPAC
has been provided by national institutions, in particular the institutions
participating in the {\it Gaia} Multilateral Agreement.

The Fermi-LAT Collaboration acknowledges generous
ongoing support from a number of agencies and institutes that
have supported both the development and the operation of
the LAT, as well as scientific data analysis. These include the
National Aeronautics and Space Administration and the
Department of Energy in the United States; the Commissariat
à l'Energie Atomique and the Centre National de la Recherche
Scientifique/Institut National de Physique Nucléaire et de
Physique des Particules in France; the Agenzia Spaziale
Italiana and the Istituto Nazionale di Fisica Nucleare in Italy;
the Ministry of Education, Culture, Sports, Science and
Technology (MEXT), High Energy Accelerator Research
Organization (KEK), and Japan Aerospace Exploration Agency
(JAXA) in Japan; and the K. A. Wallenberg Foundation, the
Swedish Research Council, and the Swedish National Space
Agency in Sweden.

This work made use of data supplied by the UK Swift Science Data Centre at the University of Leicester.

The Digitized Sky Survey was produced at the Space Telescope Science Institute under U.S. Government grant NAG W-2166. The images of these surveys are based on photographic data obtained using the Oschin Schmidt Telescope on Palomar Mountain and the UK Schmidt Telescope. The plates were processed into the present compressed digital form with the permission of these institutions.

\small
\bibliographystyle{aasjournal}
\bibliography{grb}

 \end{document}